# GRAPHIC: GatheR-And-Process in Highly parallel with In-SSD Compression Architecture in Very Large-Scale Graph

Yiming Chen, Guohao Dai, Mufeng Zhou, Mingyen Lee, Nagadastagiri Challapalle, Guodong Yin, Zekun Yang, Yongpan Liu, Huazhong Yang, Vijaykrishnan Narayanan and Xueqing Li

BNRist/ICFC, Electronic Engineering Department, Tsinghua University, Beijing, China

Email: xueqingli@tsinghua.edu.cn

## ABSTRACT

Graph convolutional network (GCN), an emerging algorithm for graph computing, has achieved promising performance in graph-structure tasks. To achieve acceleration for data-intensive and sparse graph computing, ASICs such as GCNAX have been proposed for efficient execution of aggregation and combination in GCN. GCNAX reducing 8x DRAM accesses compared with previous efforts. However, as graphs have reached terabytes in size, off-chip data movement from SSD to DRAM becomes a serious latency bottleneck.

This paper proposes Compressive Graph Transmission (CGTrans), which performs the aggregation in SSD to dramatically relieves the transfer latency bottleneck due to SSD loading compared to CMOS-based graph accelerator ASICs. In-SSD computing technique is required for CGTrans. Recently, Insider was proposed as a near-SSD processing system computing by integrating FPGA in SSD. However, the Insider still suffers low area efficiency, which will limit the performance of CGTrans.

The recently proposed Fully Concurrent Access Technique (FAST) is utilized. FAST-GAS, as an in-SSD graph computing accelerator, is proposed to provide high-concurrent gather-and-scatter operations to overcome the area efficiency problem. We proposed the GRAPHIC system containing CGTrans dataflow deployed on FAST-GAS. Experiments show CGTrans reduces SSD loading by a factor of 50x, while GRAPHIC achieves 3.6x, and 2.4x speedup on average over GCNAX and CGTrans on Insider, respectively.

## 1 INTRODUCTION

Deep neural network (DNN) has been successfully used in many artificial intelligence (AI) applications, such as object detection, voice recognition, and text classification [1][2][3]. Unlike dense image and text data, graph response the relationship across objects, which is an emerging application in e-commerce and advertisement recommendation systems [4][5].

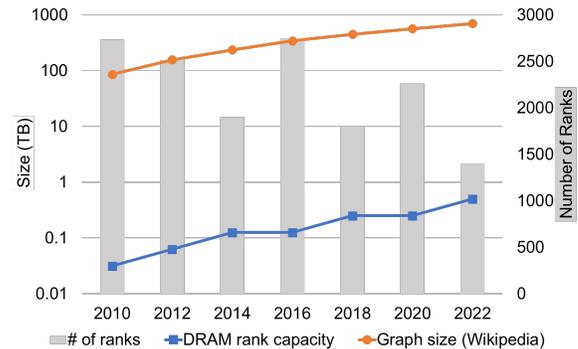

**Figure 1: Challenges in the growth of graph size and DRAM capacity bottleneck in a large-scale graph.**

However, DNN only demonstrates limited performance when directly applied to the graph data structure.

Graph computing methods, including conventional algorithms [6][7][8] and the emerging graph convolutional network (GCN) [5][9], perform well in the graph structure. Compared with DNN, which uses regular data access, graph computing requires irregular data access as a new challenge [10][11]. The emerging GCN [9] combines traditional graph computation algorithms with DNNS. An aggregation operation is first performed on all vertices/edges. It is based on the conventional graph algorithm. Then the aggregated vertices/edges are fed into a multilayer perceptron (MLP) [12] for feature extraction. GCN has achieved good performance in recommendation algorithms. However, the aggregation process of GCN has a large number of irregular visitation due to the aggregation of all neighbors of a vertex/edge. Subsequently, the proposed GraphSAGE [5] solves this problem by sampling the same number of neighbors for one vertex/edge at a time. The graphs method reduces the number of neighboring and improves the performance. At the same time, it makes the number of aggregations per-vertex/edge the same, allowing the access load to be balanced. Nevertheless, GCN accesses during computation are still highly random. Traditional DNN accelerators [13][14][15] do not apply to random access due to the sparse vertex connectivity in the graph.

Recent graph computing accelerators [16][17][18] fetch features of edges and vertices from DRAM. Aggregation is performed in the aggregation engine, while dense computing such as MLP inference will be performed in the combination engine. This approach has been thoroughly explored. The graph computing accelerator represented by GCNAX [16] has obtained high access efficiency and computational performance by eliminating redundant DRAM accesses through rational data flow scheduling. However, the amount of realistic-level graph data cannot be fully deployed in DRAM, as shown in Figure 1. There is a gap between the size of social networks and the maximum capacity of DRAM on a single rank. Thousands of DRAM ranks are needed to store the data of the entire graph. It is not possible in a single computer system. A distributed approach [19] to computing is possible. But the network transmission instability, limited speed, and scheduling difficulties greatly limit the development of graph computing. Another approach is to use SSD storage [20] for data that cannot be stored in DRAM, as shown in Figure 2 (a).

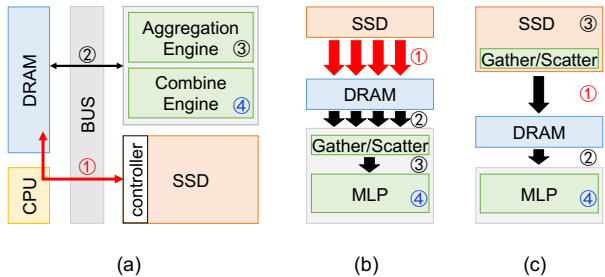

**Figure 2: (a) Current graph computing systems and (b) the challenges in SSD loading. (c) Proposed CGTrans by aggregating in SSD to save SSD loading bandwidth.**

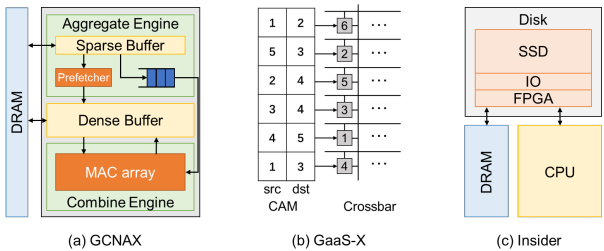

**Figure 3: Previous graph computing architectures [16][21][22].**

Figure 2 (b) shows the challenges introduced by SSDs in the graph system [23]. The introduction of SSDs dramatically reduces the efficiency of graph computing because the transfer rate of the off-chip bus is still limited. This makes SSD access a dominant bottleneck, limiting further development of graph computation accelerators. Near SSD computing [22][24] brings a new chance to address a large amount of data. However, the area efficiency is limited. It is hard to perform a complete graph neural network within SSDs.

Proposed compressive graph transmission (CGTrans) method shows in Figure 2 (c). We observe that GCN consists of two parts of computing with different characteristics. The first part is the aggregation operation. This part of the computation is relatively simple. But it needs high frequent random access. The other part is the combination. It is the same as conventional DNN computing such as MLP. The data flow and the memory access are regular, with good speedups in the systolic array [13]. Compared to the conventional graph computing schemes where aggregation and combination are executed in the same place (such as GCNAX [16]), we proposed deploying these two operations with different features on the near-SSD and side ASIC accelerator side, respectively. By separating the random large capacity data-intensive operations from the regular small-capacity data-intensive operations, the CGTrans approach reduces the amount of data transferred from SSD to DRAM (or directly to the ASIC accelerator).

The CGTrans method requires SSD with near-storage computing abilities. The previous schemes take advantage of the reconfigurability of FPGA to perform computation inside the SSD. However, these schemes perform aggregation with limited area efficiency, which reduces the efficiency of CGTrans. We construct GRAPHIC, an accelerator architecture for graph computing, using the recently proposed FAST SRAM [25]. FAST SRAM can perform independent parallel computing in-situ. Combining with the content-address memory (CAM) [26], the GRAPHIC architecture can be integrated as a cache in SSD, which can perform aggregation in GCN. Experiments show that GRAPHIC is 2x~4x faster than architecture using GCNAX and SSD. GRAPHIC achieves 5x area efficiency improvement on the aggregation task than Insider, which uses FPGA as computing units.

The major contributions of this work are listed as follows:
- Proposal of compressive graph transmission (CGTrans) to reduce the overhead of data transfer from SSD.
- Proposal of a gather-and-scatter (GAS) engine. Methods to deploy the corresponding algorithm are also presented.
- End-to-end evaluations that show advantages of CGTrans and GAS combined GRAPHIC system.

## 2 BACKGROUND

### 2.1 Graph applications

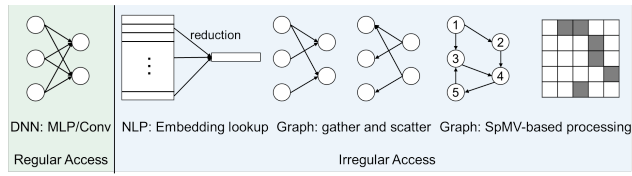

**Figure 4: A few emerging data-intensive applications**

Data-intensive applications bring serious memory wall problems due to a small amount of computation performed per memory access. There are two categories of data-intensive applications, including regular access type and irregular access type (Figure 4). The regular access type is typically represented by DNN [27]. The main manifestations are regular weight access, alignment calculation, and dense matrix representation. Irregular access type is gradually being applied in the emerging field of artificial intelligence. Examples include feature lookup and embedding for natural language processing (NLP) [28] and recommendation system [4], involving the gather-and-scatter method in the graph, and graph processing based on sparse matrix and vector (SpMV) [6][7]. Their common feature is that the computation is simple, but the operands are sparsely stored in memory. The irregular and large access requirements create difficulties in hardware acceleration, which is the key to the design.

GRAPHIC

Graph Convolutional Network (GCN) is an emerging graph algorithm used in advertisement recommendation and branch prediction systems. Each layer of GCN could be divided into two steps, which become aggregation and combination. In the aggregation step, each vertex of the graph obtains information from its neighbors. An aggregation function turns this information into a new vertex feature, such as max or sum. After the aggregation step, the graph will be input into the MLP as a dense feature, called the combination step. The MLP will further learn the features in the graph. A combination step will transfer the aggregated features to the new feature space. In summary, GCN contains both regular access and irregular access computing modes.

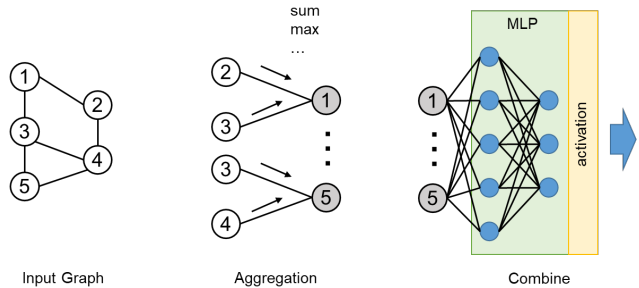

**Figure 5: Graph Convolutional Network (GCN) [9].**

The problem with GCNs is that the number of neighbors of a vertex could be huge, which will become the bottleneck of the aggregation step. At the same time, the number of neighbors of a point tends to be unbalanced. It can also lead to unbalanced memory/SSD accesses. GraphSAGE [5] was proposed to solve the problem of too many neighbors. A fixed number of neighbors will be sampled from all neighbors. This reduces the amount of computation and ensures load balancing.

## 2.2 FAST SRAM

The von Neumann architecture separates computation and storage. It leads to significant overhead in data-intensive applications. In this case, in-memory computing was proposed to overcome problems due to data movement. This architecture places computing units close to the data memory cells. The in-memory computing (IMC) method reduces data movement, improving energy efficiency in data-intensive applications.

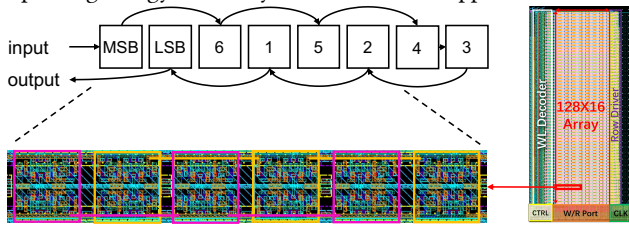

**Figure 6: Array layout of FAST SRAM [25], and an example of 8-bit cyclic shifting.**

However, one of the IMC methods, the crossbar-based solution [31], can only perform matrix-vector multiplication. While custom SA-based near-memory computing [32] can only perform bit-wise computation between two rows, limiting the throughput in highly parallel applications. A new IMC method that adapts to multibit highly concurrent computing is needed.

A Fully Concurrent Access Technique (FAST) [25] was recently proposed to support independent row-level parallelism computing (Figure 6). A shiftable SRAM design enables row-level parallelism. A controlled inverters loop achieves the cyclic shift of the bits of all words in the SRAM in each SRAM cell. Further, the feature that each row can be independently shifted in parallel is used to achieve independent parallel computation for each row. A 1-bit ALU is attached for each row. It accepts the input from the last cell of the row and writes the output result to the first cell. By combining the cyclic shift and the 1-bit arithmetic operation, row-level multibit operations could be performed independently in parallel, leading to a novel method of in-memory computing. Putting computation and storage units together eliminates the overhead of SAs and reduces the cost of data moving. The 1-bit calculation uses a sequential bit-serial algorithm [33] with only a 1-bit processing unit to complete the multi-bit operation (such as *add*, *mul*), which saves area in high-parallelism computing. This scheme could support applications that need parallel updates to the stored data in several rows by replacing ALUs with different functions.

## 2.3 Graph computing accelerators

Recently, some works [16][17][18] have presented well-performance architectures for graph computing. GCNAX [16] represents typical fully-digital graph computing architectures. These architectures have similar characteristics. The accelerator could be divided into an aggregation engine and a combination engine. These two parts have different data flows, which adapt to the calculation method of different features in graph computing. Sparse graph connections are stored in the data buffer in the aggregate engine. The aggregate engine is mainly responsible for sampling neighbors and aggregating them. The aggregated vertices will become a dense format conducive to MLP. Finally, the features stored in a dense format buffer will be fed to the MAC array in the combination engine. GCNAX achieves excellent performance by optimizing data flow, leading to a balanced load on computing units and no redundant DRAM accesses.

Computing-in-memory is a technique that has received attention and been intensively studied in CNN inference. Recently, GaaS-X [21] was presented as an RRAM-based computing-in-memory architecture for graph computing. It uses the coordinate (COO) format to store the sparse matrix in content-addressable memory (CAM) and crossbar. CAM [26][34] is a kind of associative memory that can support in-situ data search without the need to load data to the external and set the output of matched row. GaaS-X uses CAM to select the rows of the crossbar to complete sparse processing. Recently, a GCN computing architecture based on this crossbar structure is also shown as PIM-GCM [35]. The architecture support node-stationary data flow, which could reduce off-chip memory access and increase the reuse of node feature data in the aggregation step.

The entire graph cannot be stored in DRAM as the data grows. Severe latency bottlenecks due to additional SSD accesses limit the performance improvement of graph computing architectures. Near-storage computing for SSD [22][24] can alleviate the lack of bandwidth for SSD access. The CPU can request read and write instructions to the SSD normally. Also, it can send requests to the FPGA, a reconfigurable module that can be designed in hardware for different applications. The FPGA processes data within the SSD, thus avoiding the overhead of transferring large amounts of data on the off-chip bus. Programming of the FPGA is done through the CPU. A complete programming model, OpenCL, is already available for storage management, while the HLS or RTL language is used for FPGA.

## 3 GRAPHIC ARCHITECTURE

We propose a compressive graph transmission technique to perform aggregation and combination separately on the near-SSD and the ASIC sides. An SRAM-cache-based gather-and-process in highly parallel with in-SSD compression (GRAPHIC) architecture is presented to execute high-area-efficiency aggregation in SSD.

### 3.1 Overall architecture

Figure 7 shows the overall architecture of the GRAPHIC system, which contains SSD, DRAM, and PE array. Where the graph data is stored in the SSD, in the aggregation step, the features of vertices/edges will be aggregated in the gather-and-scatter (GAS) cache. The GAS cache consists of CAM and FAST-SRAM. CAM is used to locate the neighbors of the selected vertex, while FAST-SRAM is used to perform parallel aggregate operations. GAS caches are used as the aggregation engine in the GRAPHIC system. The combination engine then fetches the aggregated features. In the GRAPHIC design, the conventional systolic arrays as PEs are supposed to serve as MLP accelerators. In the combination step, the features will be transformed into fully-connected layers. After the calculation, new features will be obtained and these features will be re-programmed back to the SSD.

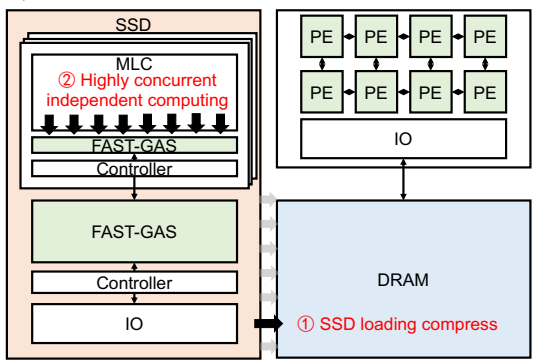

Figure 7: Overall architecture of the GRAPHIC system.

### 3.2 Compressive graph transmission

Figure 8 (a) shows the conventional data flow in previous graph computing accelerators. GCN will use all vertices and edges in the graph, which is unacceptable in very large-scale graph computing. GraphSAGE was proposed to perform inference by sampling the points. Since conventional SSDs cannot perform computation, the sampled vertices and their neighbors have to be transferred to a graph computing accelerator for aggregation, which will suffer from the slow off-chip SSD bus. Figure 9 shows the timing control of the conventional system using an SSD to store a very large-scale graph and an ASIC chip with DRAM to perform graph convolutional networks. With the introduction of SSDs into graph computing systems, it will become a new bottleneck.

Given the high transfer overhead, is it possible to compress the data on the bus?

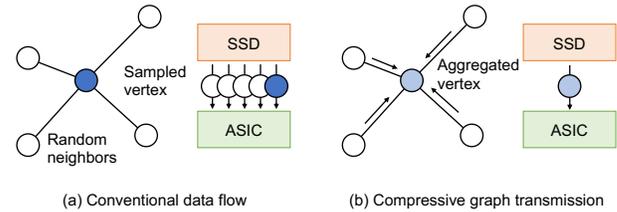

Figure 8: (a) conventional data flow transfers all features needed and (b) proposed compressive graph transmission.

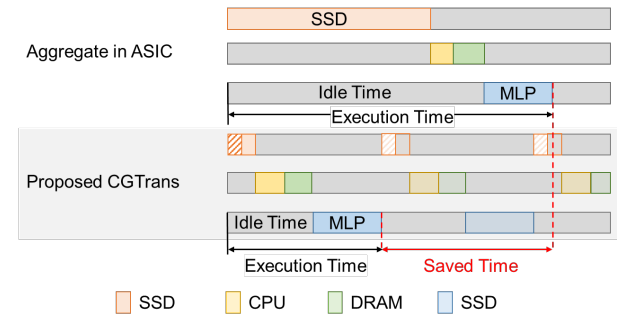

Figure 9: Latency optimization of the proposed CGTrans compared with the conventional dataflow.

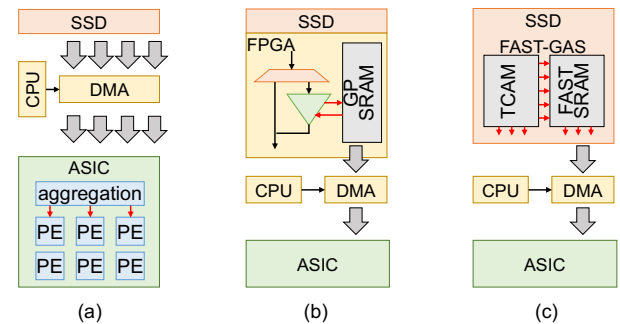

Figure 10: (a) Conventional execution requires huge loading from SSDs, while CGTrans by: (b) Near-SSD and (c) GAS-engine perform aggregation to save loading overhead.

We propose the compressive graph transmission (CGTrans) technique to overcome the disadvantages of slow SSD. Figure 8 (b) shows that this method performs the aggregation in the SSD and transfers the aggregated features to the dense computing engine. It can exponentially reduce the transfer overhead of the SSD, thus improving system performance. As shown in Figure 9, comparing the timing sequence, CGTrans will gain a significant speedup from the reduced amount of data transferred by the off-chip bus.

We observe that DRAM can store all the weights and features of a fully-connected layer during the combination step. Meanwhile, the FLOPS required for the MLP are much larger than the simple aggregation. Thus, performing MLP inference near SSD can be very costly, which will drop the performance of the graph computing system.

Figure 10 shows the execution flows of the baseline and the proposed compressive graph transmission technique. The thickness of the arrows in the figure represents the amount of transferred data on the corresponding bus. Figure 10 (a) shows the baseline, i.e., the system scheme based on GCNAX. After the vertices are moved from the SSD by the off-chip buffer, they are aggregated and combined in the GCNAX. Figure 10 (b) presents a scheme based on near-SSD computing to implement a compressive graph transmission technique. Near-SSD computing relies on FPGAs to perform aggregation within the disk. This scheme using CGTrans can reduce the SSD transfer latency. However, the area efficiency of this solution is low, making the aggregation process itself a bottleneck. Figure 10 (c) proposes a gather-and-scatter (GAS) engine consisting of FAST SRAM and CAM. Leveraging the row concurrency of FAST SRAM and CAM, the FAST-GAS shows better performance in aggregation tasks. At the same time, the 1-bit-based bit-serial algorithm provides GAS with high area efficiency. The features aggregated by the FAST-GAS are transmitted via the SSD bus to the PE array, which performs high-speed neural network inference on the chip.

## 3.3 FAST-based gather-and-scatter (FAST-GAS)

The FAST SRAM introduced in section 2 can perform an independent cyclic shift in each row. With the help of a 1-bit ALU attached behind each row, which can memory 1-bit information, multi-bit arithmetic operations are possible after one cyclic shifting. Also motivated by GaaS-X [21], combining the independent cyclic shift with the parallel lookup capability of content-address memory (CAM), a data-dependent high-parallel computation can be implemented. Based on such a new logic of find-and-computing, we propose the gather-and-scatter (GAS) engine (Figure 11 (a)).

CAM is a memory architecture that enables high-speed parallel lookup. In general, it contains an array of memory cells, an input driver, and a matching decoder. A CAM accepts input that is waiting to be matched in the memory array, while each row has a matching line (ML) for outputting the result of a match or not. The address of the matched value is output through the priority decoder. It was initially designed for IP address matching and forwarding. It is also used for irregular data selection in the GaaS-X architecture. However, the decoder is also the most significant bottleneck when using CAM directly in the computing system. One difficulty applying CAM's highly concurrent search to computing is that the matching results are not easily applied to parallel computation units. First, if there are a lot of rows in the CAM array, the area, and energy overhead of the decoder will be very large. Then, the decoder can output only one matching address at a time.

When multiple rows are matched in an array, it is difficult to output the results altogether. This can lead to a decrease in the performance of CAM in highly concurrent computations. Direct use of fully-digital FIFOs that store all weights and ALUs architecture would introduce the significant area and power overhead. Even though CiM-based processing could be an alternative by activating the matching rows simultaneously, this approach would require considerable overhead components such as the ADC and digital peripherals. Besides, due to data sparsity, each ADC will only perform an average of 4-row addition in parallel on a real-world dataset which further lowers energy efficiency.

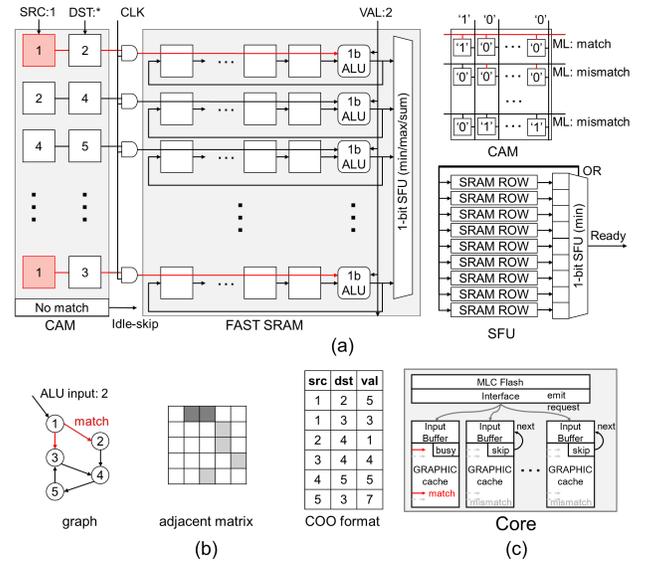

Figure 11: Proposed GRAPHIC cache: (a) structure overview shown with (b) a sample graph, and (c) system architecture with idle-skip capability in SSD.

The gather-and-scatter (GAS) engine innovatively utilizes the advantages of both CAM and FAST SRAM. It combines CAM and FAST SRAM to achieve high-efficiency irregular access and processing. The main idea is to discard the area and power costly decoder in CAM and use the matched line output of CAM directly. Thanks to the independent parallel computation characteristics of FAST, the match lines of CAM can control each row independently of the SRAM. Specifically, in the GRAPHIC architecture for graph computing, the CAM stores the source and destination vertices of each edge, whose features are stored in the same row of FAST SRAM. The sample in Figure 11 (b) shows that vertex 1 is matched in the CAM as the source vertex of the stored edge. Then two matching outputs representing weights of two edges (shown in red) will activate the clock of FAST SRAM in the corresponding row, while the 1-bit ALU performs weight computation and update in SRAM. Considering the high sparsity of the graph, only clock-activated rows are shifted. Other deactivated rows remain stationary to avoid energy waste in the useless cyclic shifts.

Compared with fully-digital FIFO-based and FPGA-based approaches, the FAST-GAS exhibits higher performance and area efficiency thanks to the low-cost in-situ computing units. Therefore, the proposed GAS cache becomes a promising alternative to the conventional CPU cache. Figure 11 (c) shows that an idle-skip strategy is used to improve the concurrency further. In detail, a buffer much smaller than SRAM is added before each GAS input which receives a matching source and destination from the global ALU. If the CAM fails to match the input, the GAS cache will skip to the next round of computation. Figure 11 (c) also shows how the FAST-GAS can be integrated into the SSD to make it support the CGTrans technique.

### 3.4 Computing methods

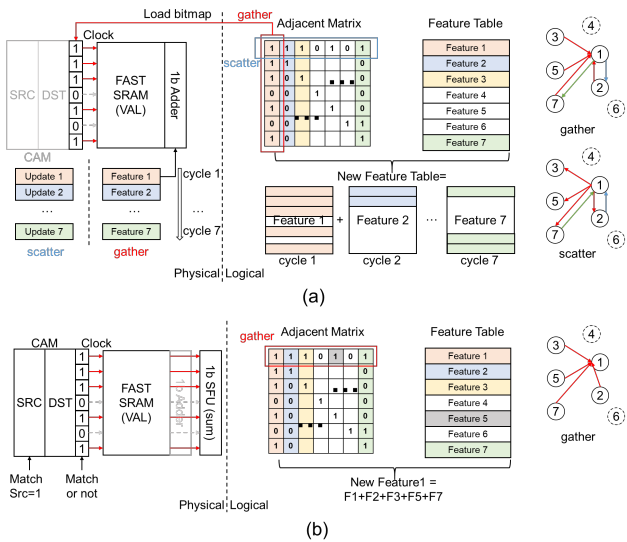

matrix will be generated and loaded as the activation of the clock of FAST SRAM in order. Note that the adjacent matrix does not need to be stored directly, as its size is always considerable. Each activated row of the FAST SRAM will perform an aggregation function in situ between the stored features and the input features corresponding to the current column. In each cycle, the feature of one vertex will be added to the features of its neighbors by FAST SRAM. The value stored in the FAST SRAM is the aggregated features.

Another approach is shown in Figure 12 (b), where the information of edges and vertices is stored in CAM and FAST SRAM in COO format (shown in Figure 11 (b)). The indexes of the origin and the destination of each edge are stored in the same row of a CAM, while the features are stored in the FAST SRAM. Unlike the previous method, the information of the vertices to be aggregated comes from the matching result from CAM. Specifically, all vertices need to be walked through. Each vertex is selected and matched as a destination in CAM. The matching CAM row activates the corresponding FAST SRAM row to perform aggregation. The matched features in FAST SRAM will be shifted and added bit by bit through FAST SRAM and the external 1-bit Special Function Unit (SFU) [20]. This method uses additional CAM and SFU units but does not require a dense bitmap generated from a sparse format graph. Different aggregate functions are implemented with bit-serial by different 1-bit ALU and SFU arithmetic functions.

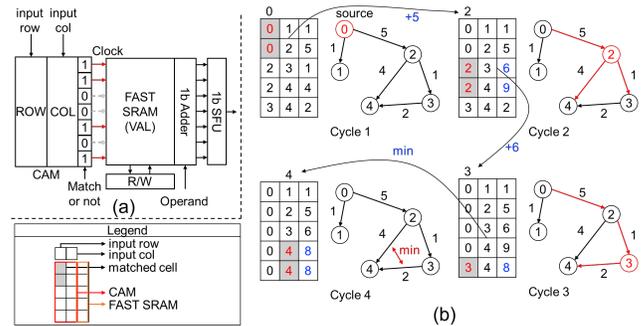

**Figure 13: GAS-based SSSP processing: (a) structure, and (b) step-by-step operations.**

**Figure 12: GAS-engine can perform (a) dense computing and (b) sparse COO format computing.**

Based on the FAST-GAS, many accelerations methods for irregular computations can be achieved in the SSD. GAS will directly replace the read/write cache in SSD to reduce area overhead. There are multiple forms of aggregation in GCN. Among them, summation and maximum are the two most commonly used aggregate functions. Figure 12 shows two ways of deploying aggregation (feature embedding) algorithms. The implementation relies on gathering performed by CAM and the process performed by FAST SRAM. Figure 12 (a) shows a gather-and-process method under the dense bitmap format graph. At the logical level, the adjacency matrix of the graph is expanded in a dense form. The right part of Figure 12 (a) shows that the row direction represents the source vertex, and the column direction represents the target vertex. In this method, '0' and '1' in the ith row of the adjacency matrix mean whether the vertex on the corresponding column needs to be aggregated with the ith vertex. The '1' in the jth column of the adjacency matrix represents the total vertices that need to be aggregated with the jth vertex. To perform aggregations, Columns of the adjacent

The above shows the method of feature embedding using the FAST-GAS. In addition to aggregation, the FAST-GAS also supports traditional graph computation algorithms. Its flexible architecture of CAM+FAST SRAM also lays the foundation for supporting new algorithms in the future.

Figure 13 presents the breadth-first search (BFS) algorithm implemented on GRAPHIC cache with in-situ computation and update. Taking the (single-source shortest path) SSSP algorithm as an example, the graph is stored in a COO format in CAM and FAST SRAM. First, neighbors (vertex 1 and vertex 2) of the source vertex (vertex 0) are matched through CAM. Since vertex 1 is terminated, vertex 2 becomes the source whose matching results are vertex 3 and 4. The edge weight of vertex 2 and vertex 0 is used as an operand of the 1-bit ALU for edge weights update, which completes the shortest path update from source vertex 0 to vertex 2. Each atomic operation consists of addition and minimization. Before the procedure, the values stored in the FAST SRAM rows are the weights of the corresponding edges. After add operation, the values stored in the rows are the weights of the paths from the starting point to the destination of the edge. The source vertex of the corresponding edge is replaced as 0 (starting vertex). Finally, the minimum weight of the same destination to the starting point is taken to get the single-source shortest path of that vertex. The atom operation will be repeated until all stored source points end up with source vertex 0. It should be noted that all procedures are finished in the proposed GRAPHIC cache with only two intermediate registers, resulting in low area and energy overhead with conventional CPU cache computation.

The connected component (CC) algorithm can be performed on the FAST-GAS in the following way. Initialize the CAM and FAST SRAM value, same as the SSSP algorithm. Traverse the vertices as follows. First, match the current vertex index as the destination in CAM and use SFU to find-and-update the minimum data among matched rows in FAST SRAM. After this atomic operation, the data in corresponding rows will become the smallest one in these rows. When the traversal is completed, the connected component lookup is performed.

When the FAST-GAS performs sorting, it uses *insert sort*. The first element to sort in each cycle is compared with each row by FAST SRAM. 1-bit ALUs serve as a compare-only function. A 1-bit flag is needed to store and output the result. Then, use SFU (1-bit adder tree) to get the sum of the flags. The result is the sorted location of the first row. Finally, switch the first row and the row of this location. If no row is hit, the current first element is sorted, and the next element serves as the first element to sort in the next cycle. Using the full concurrency feature of FAST SRAM, the $O(n^2)$ *insert sort* is optimized to $O(n)$, surpassing the *quick sort*. This can also be used as a sub-task for broken-down divide-and-conquer algorithms (such as quick sort) placed in FAST SRAM before execution when dealing with vast arrays. It can also be used in cases where only the order of several numbers in an array is required.

## 4 IMPLEMENTATION RESULTS

### 4.1 Evaluation of FAST-GAS

**Table I: Hardware evaluation settings**

| Process (nm) | 65 | |
|---|---|---|
| Component | FAST SRAM | CAM |
| Array Size | 128x16 | 128x16 |
| Area (mm²) | 0.016 | 0.013 |
| Energy (pJ) | 0.38 / OP | 0.33 / OP |
| Latency (ns) | 0.025 / OP | 0.182 / OP |

* OP: 16-bit addition with data written back

The FAST-GAS has the advantage of higher area efficiency compared to the digital solutions approach and Insider. The area efficiency of digital solution and FPGA is evaluated in Vivado and DC synthesis. The performance of the FAST-GAS will be evaluated by post-layout SPICE simulation in Cadence. The configuration is shown in Table I.

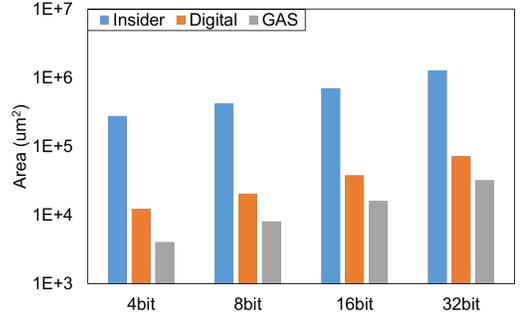

**Figure 14: Area comparison among Digital, Insider, and proposed GAS-engine under the same throughput.**

Figure 14 compares the area efficiency of the FAST-GAS, the digital circuit, and the Insider. It can be seen that replacing the cache in SSD with a FAST-GAS gives high performance at a small area cost. The advantages of area efficiency come from the fully parallel independent computation achieved by the in-memory shift structure of the FAST SRAM cell and CAM matching. It also solves the difficulty in utilizing CAM high concurrent matching data by removing the matching decoder.

### 4.2 Evaluation of CGTrans

Based on the circuit evaluation in SPICE, we evaluate the reality-level graph shown in Table II. GCNAX, as a state-of-art GCN AISC is selected as the baseline. The system is deployed in a classic computer system containing a CPU to schedule, several DRAMs, a solid-state disk, and an accelerator ASIC. GCNAX is set as the ASIC accelerator in the baseline, which serves as an aggregation and combination engine. The graph is too large to be stored entirely in DRAM. SSD is introduced as data storage. In the proposed CGTrans program, a general-purpose neural network accelerator (systolic array) is used as a combination engine. The SSD is GRAPHIC architecture. The SRAM cache is replaced as the FAST-GAS to perform aggregation in SSD.

We have performed application-level simulations on the GRAPHIC system using a custom simulator developed with networkX [36] and PyTorch by tracing data and operations. Input graphs are stored in standard COO format. The performance differences due to different dataset sizes are taken into account. This work models the difference between the proposed CGTrans and the data flow in previous DRAM-based GCNAX. The rest of the computation overheads are identical to that in [23]. The algorithm deployed on the system is GraphSAGE. Taking the first layer as an example, GraphSAGE samples 50 neighbors at a time according to the general setup.

**Table II: Dataset Configuration**

| Name | Nodes (M) | Edges (B) | # of Features |
|---|---|---|---|
| Reddit | 37.3 | 53.9 | 602 |
| Movielens | 22.2 | 59.2 | 1000 |
| Amazon | 265.9 | 9.5 | 32 |
| OGBN-100M | 179.1 | 5.0 | 32 |
| Protein-PI | 9.1 | 8.8 | 512 |

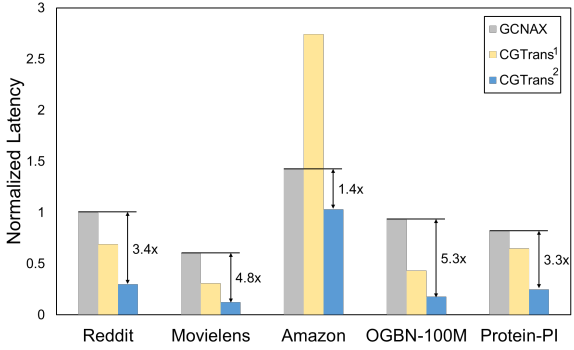

*1: CGTrans with a Near-SSD scheme
*2: CGTrans with the proposed FAST-GAS

**Figure 15: Normalized latency comparison of GCNAX and CGTrans data flow.**

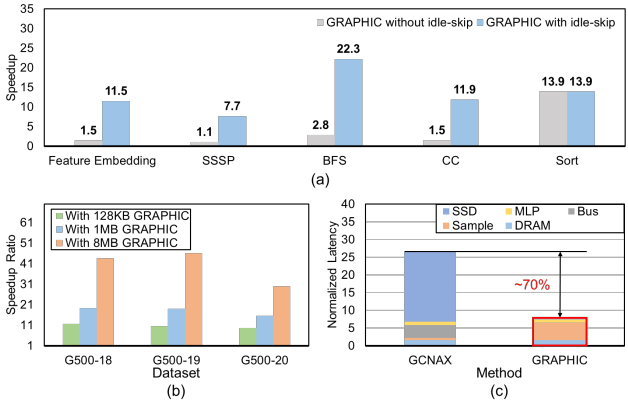

**Figure 16: Benchmark results in (a) various algorithms, (b) different dataset scales and cache sizes, and (c) GCN.**

Figure 15 compares the latency of GCNAX and CGTrans data flow in graphs of different sizes and connect relationships. CGTrans, based on the Near-SSD scheme, could achieve about 50% latency reduction, except for Amazon. The gain in this approach is limited due to the overhead of aggregation in FPGA. Though SSD accesses are much reduced, the aggregation step becomes a new bottleneck. Under the same area, the throughput in near-SSD is limited by inefficient FPGA or digital synthesis. CGTrans, with the proposed FAST-GAS, overcomes this issue. The high concurrency of both CAM and FAST SRAM brings beyond digital performance to the FAST-GAS. The proposed GRAPHIC system integrated with CGTrans and FAST-GAS achieves 0.4~4.3x performance improvement in GraphSAGE applications.

### 4.3 Evaluation of GRAPHIC system

Figure 16 (a) shows the performance of various graph computing algorithms on the GRAPHIC system with a 1MB FAST-GAS. Feature Embedding and BFS are deployed for wiki-Talk. SSSP task is deployed for road-Italy-osm. CC is set up on the Wikipedia-2009 dataset. GAS without idle-skip gains only about 0.4~1x times speed-up in aggregation and conventional graph algorithms due to the sparsity and unbalanced load in most graphs. The speed-up ratio is limited in pure GAS because the FAST SRAM will primarily be on standby in this case. The proposed idle-skip strategy in GAS further explores the inherent parallelism of the FAST-GAS and achieves an average speed-up of 10.1x compared with a typical cache. There is also a 2.7x~3.1x improvement over near-SSD [22][24] solutions

Figure 16 (b) shows the performance comparison of running the BFS on the G500 dataset at different scales and in various sizes of the FAST-GASs of the GRAPHIC system. The effectiveness of the FAST-GAS is also related to the size of datasets and cache. With the vertex-orientated graph partitioning method, the speedup ratio will increase as the size of the cache increases. Thanks to idle-skip, the idle rate of FAST SRAM is significantly reduced. In this way, acceleration is still guaranteed when the cache is smaller than the graph size.

Figure 16 (c) shows the system-level end-to-end evaluation using aggregation and combination in GCN as an example. The Reddit dataset shown in Table II is used in the evaluation. Overall, the GRAPHIC system gains about a 70% reduction in latency compared to GCNAX. The detailed latency breakdown is presented in the bar. Although the aggregation calculation within the SSD is slower than that of the combination engine in the ASIC, it can be found that the CGTrans technique used in GRAPHIC dramatically reduces the latency overhead of SSDs in off-chip transfers compared to GCNAX. The significant reduction in SSD transfer latency improves the overall performance of the system.

On average, experiments show CGTrans reduces SSD loading by a factor of 50x, while GRAPHIC achieves 3.6x, and 2.4x speedup over GCNAX and CGTrans on Insider, respectively.

*Future work.* GAS is an attempt to take on the independent concurrency offered by FAST SRAM. Further exploration is still worth analyzing. Meanwhile, in-Flash computing and direct read-out of aggregation results will perform better on CGTrans.

## 5 Conclusion

This paper proposes a compressive graph transmission (CGTrans) to reduce load overhead from SSD in very large-scale graph computing systems. An implementation of the high area efficiency GRAPHIC system with CGTrans to compute in SSD is also presented. First, the bottleneck of aggregation in very large-scale graph convolution networks is observed. The key idea of the CGTrans technique is to perform aggregation computing within SSD to compress the feature data on the SSD bus. Meanwhile, by combining CAM and FAST SRAM, graph data process can be completed in the area-efficient gather-and-scatter (GAS) engine. By avoiding high-overhead retrieval and access, the FAST-GAS takes advantage of high concurrent matching in CAM to accomplish neighbor sampling in graph computation and high concurrent computing in FAST SRAM to perform features aggregation and conventional graph algorithm. The proposed GRAPHIC system consists of the CGTrans technique and the FAST-GAS. We validate the proposed design in various algorithms and datasets. Evaluations show that the proposed CGTrans technique gains an average of 2.6x improvement in GCN applications compared with the previous GCNAX.

GRAPHIC